\documentclass[epj,final]{svjour}
\usepackage{graphicx}  
\usepackage{amsmath,amssymb}  
\usepackage{cite}  
\begin{document}
\hyphenation{To-mi-na-ri}
%
\title{
Smearing origin of zero-bias conductance peak \\ 
in Ag-SiO-Bi$\vec{_\mathsf{2}}$Sr$\vec{_\mathsf{2}}$CaCu$\vec{_\mathsf{2}}$O$\vec{_\mathsf{8+\delta}}$
planar tunnel junctions: \\
influence of diffusive normal metal verified with the circuit theory
}
\author{Iduru Shigeta\inst{1,}\thanks{\emph{Present address:} Department
of Physics, Kagoshima University, Kagoshima 890-0065, Japan \vfill 
\email{shigeta@sci.kagoshima-u.ac.jp}},
Yukio Tanaka\inst{2,3}, 
Fusao Ichikawa\inst{4},
\and Yasuhiro Asano\inst{5} 
}                     
%
%
\titlerunning{Smearing origin of zero-bias conductance peak in
Ag-SiO-Bi-2212 planar tunnel junctions}
\authorrunning{I. Shigeta \textit{et al}.}
\institute{Department of General Education, Kumamoto National College of
Technology, Kumamoto 861-1102, Japan
\and
Department of Applied Physics, Nagoya University, Nagoya 464-8603, Japan
\and
CREST, Japan Science and Technology Agency (JST), Nagoya 464-8603, Japan
\and
Department of Physics, Kumamoto University, Kumamoto 860-8555, Japan
\and
Department of Applied Physics, Hokkaido University, Sapporo 060-8628,
Japan
}
\date{Received 28 January 2006 / Received in final form  23 October 2006}
%
\abstract{
We propose a new approach of smearing origins of a zero-bias
conductance peak (ZBCP) in high-$T_\mathrm{c}$ superconductor tunnel
junctions through the analysis based on the circuit theory for a
$d$-wave pairing symmetry.
The circuit theory has been recently developed from conventional
superconductors to unconventional superconductors.
The ZBCP frequently appears in line shapes for this theory, in which the
total resistance was constructed by taking account of the effects
between a $d$-wave superconductor and a diffusive normal metal (DN) at a
junction interface, including the midgap Andreev resonant states (MARS),
the coherent Andreev reflection (CAR) and the proximity effect.
Therefore, we have analyzed experimental spectra with the ZBCP of 
Ag-SiO-Bi$_{2}$Sr$_{2}$CaCu$_{2}$O$_{8+\delta}$ (Bi-2212) planar tunnel
junctions for the \{110\}-oriented direction by using a simplified
formula of the circuit theory for $d$-wave superconductors.
The fitting results reveal that the spectral features of the ZBCP are
well explained by the circuit theory not only excluding the Dynes's
broadening factor but also considering only the MARS and the DN
resistance.
Thus, the ZBCP behaviors are understood to be consistent with those of
recent studies on the circuit theory extended to the systems containing
$d$-wave superconductor tunnel junctions.
\PACS{
       {74.25.Fy}{Transport properties (electric and thermal
       conductivity, thermoelectric effects, etc.)} \and
       {74.45.+c}{Proximity effects; Andreev effect; SN and SNS
       junctions} \and
       {74.50.+r}{Tunneling phenomena; point contacts, weak links,
       Josephson effects} \and
       {74.72.Hs}{Bi-based cuprates}
     } 
} 
\maketitle
\section{Introduction}
\label{sec:introduction}
The experimental studies of zero-bias conductance peak (ZBCP) behaviors
have been frequently reported for various unconventional superconductors
with an anisotropic pairing symmetry, for instance, $p$-wave and $d$-wave
pairing symmetries~\cite{Lesu92,Tair98,Wei98,Kash00,Suzu01,Mao01,Shar01,Lofw01,Shig02,Frea03a,Mao03,Qazi03,Frea03b,Miya03,Kash04,Kawa05,Deut05}.
The ZBCP that results from the Anderson-Appelbaum scattering 
has been studied thoroughly via previous
experiments~\cite{Appe66,Ande66,Appe67,Sand94,Covi96}.
However, only until recently was there an alternate theory to explain
the origin of the ZBCP.
In this theory, the ZBCP has been understood as the formation of the
midgap Andreev resonant states (MARS) at a junction interface in a ballistic
normal metal-insulator-superconductor (N/I/S) tunnel
junction~\cite{Tana95}.
A basic theory of ballistic transport in the presence of the MARS has
been formulated in the case of $d$-wave superconductors by way of the
Blonder-Tinkham-Klapwijk (BTK) theory~\cite{Blon82}.
This model has sufficiently explained ZBCP behaviors in
high-$T_\mathrm{c}$ cuprate superconductors.
Another mechanism of the ZBCP plays a role in the coherent Andreev
reflection (CAR), which induces the proximity effect in the diffusive
normal metal (DN)~\cite{Volk93} at a diffusive normal
metal-insulator-superconductor (DN/ I/S) junction interface.
Then, there are two kinds of the ZBCP due to the formation of the MARS at 
junction interfaces of $d$-wave superconductors and of that due to the CAR
by the proximity effect in the DN.

\begin{figure}[tb]
\centerline{\includegraphics*[bb=30 115 550 505,width=8.3cm,keepaspectratio,clip]{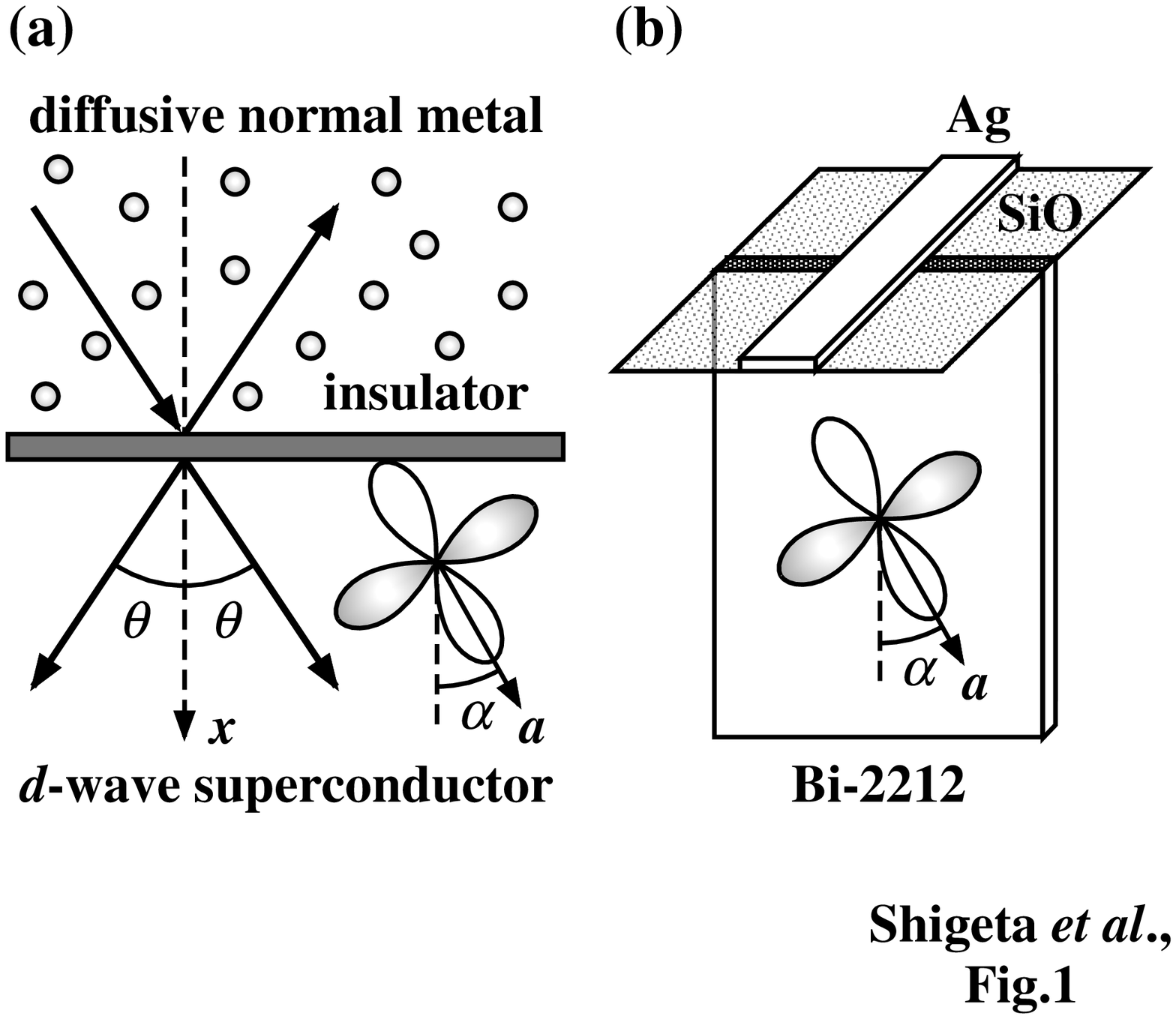}}
\caption{(a) Schematic illustration of incoming and outgoing processes
 at the junction interface for the diffusive normal
 metal-insulator-unconventional superconductor tunnel junction.
 The unconventional superconductor junctions can be incorporated into the
 circuit theory by means of the relation between matrix current and
 asymptotic Green's functions~\cite{Tana03a}.
 This relation accounts for anisotropic features of the $d$-wave
 superconductors, as sketched for a $d$-wave superconductor.
 (b) Planar type junction layout for the $ab$-plane direction.
 The Ag-SiO-Bi-2212 planar tunnel junction contains the SiO layer
 between the Ag thin film of the counterelectrode and the Bi-2212 single
 crystal enclosed with epoxy resin.
 The SiO barrier and Ag counterelectrode are deposited perpendicular to 
 the $ab$-plane of Bi-2212 single crystals.
\label{fig:1}}
\end{figure}

The circuit theory includes both effects of the MARS and the CAR due to
constructing the theory at a junction interface under the condition of
DN/I/S tunnel junctions for $s$-wave, $p$-wave and $d$-wave
superconductors~\cite{Naza94,Naza99,Tana03a,Tana03b,Tana04a,Tana04b,Yoko05a,Tana05,Yoko05b}.
Figure~\ref{fig:1}a illustrates the schematic drawing at a junction
interface of a $d_{x^2 - y^2}$-wave superconductor for the circuit
theory.
In this theory, the ZBCP frequently appears in the line shapes of
normalized tunneling conductance $\sigma_\mathrm{T}(eV)$, and
we always expect the ZBCP to be independent of $\alpha$ for low
transparent junctions with the small Thouless energy $E_\mathrm{Th}$.
Here, $\sigma_\mathrm{T}(eV)$ is defined as dividing tunneling
conductance $\sigma_\mathrm{S}(eV)$ in the superconducting state by
tunneling conductance $\sigma_\mathrm{N}(eV)$ in the normal state, and
$\alpha$ denotes the angle between the normal to the interface and the
crystal axis of $d$-wave superconductors.
The nature of the ZBCP due to the MARS and that due to the CAR are
significantly different.
The corresponding $\sigma_\mathrm{T}(0)$ for the former case can take
arbitrary values exceeding unity.
On the other hand, $\sigma_\mathrm{T}(0)$ for the latter case never
exceeds unity.
Furthermore, the ZBCP width in the former case is determined by
the transparency of the junction, while the width in the latter case is
determined by $E_\mathrm{Th}$. 
These two ZBCPs compete with each other since the proximity effect and
the existence of the MARS are incompatible in singlet junctions.

The influence of the resistance $R_\mathrm{d}$ in a DN is significant
for the resulting $\sigma_\mathrm{T}(eV)$.
Hence, for the actual quantitative comparison with tunneling
experiments, we must take into account the effect of $R_\mathrm{d}$.
In the circuit theory for $d$-wave superconductors, the ZBCP is
frequently seen in the line shapes of $\sigma_\mathrm{T}(eV)$.
For $\alpha = 0$, the ZBCP is due to the CAR.
However, for $\alpha \neq 0$, the robustness of the ZBCP does not depend
on the DN resistance $R_\mathrm{d}$. 
In such an extreme case as $\alpha = \pi/4$, the ZBCP arises from the
MARS, and the CAR and the proximity effect are absent.
The $\sigma_\mathrm{T}(eV)$ is then given by an elementary application
of Ohm's law:
$\sigma_\mathrm{T}(eV) = (R_\mathrm{b} + R_\mathrm{d}) /
(R_{R_\mathrm{d}=0} + R_\mathrm{d})$~\cite{Tana03a,Tana04a}, where
$R_\mathrm{b}$ is the resistance from the insulating barrier in a normal
state and $R_{R_\mathrm{d}=0}$ is the total resistance of the DN/I/S
tunnel junction in the condition of $R_\mathrm{d} = 0$.
In the case of $R_\mathrm{d} = 0$, there is no proximity effect in the
DN, and the junction resistances are given by the quasiballistic
formulas of reference~\cite{Tana95}.
Therefore, the theoretical results serve as an important guide to
analyzing the actual experimental data of the tunneling spectra of
high-$T_\mathrm{c}$ cuprate junctions.

In the present paper, we will discuss the ZBCP behaviors in
Ag-SiO-Bi$_{2}$Sr$_{2}$CaCu$_{2}$O$_{8+\delta}$ (Bi-2212) planar tunnel
junctions 
by using the circuit theory for $d_{x^{2} - y^{2}}$-wave superconductors
in the case of $\alpha = \pi/4$, and then,  for analysis of the ZBCP, we
will focus on the MARS case involving an influence of $R_\mathrm{d}$,
not the CAR or the proximity effect cases.

\section{Experimental details}
\label{sec:experimental}
Planar type tunnel junctions are fabricated in some experiments, while
other measurements rely on scanning tunneling microscopy and
spectroscopy (STM/STS) for recent spectroscopic purposes.
The STM/STS has the advantages of versatility adjustable junction
resistance, simultaneous topographical imaging at atomic scale resolutions
together with the spatial variations of local density of states (LDOS),
and momentum space ($\vec{k}$-space) images obtained by Fourier
transformation of real space ($\vec{r}$-space)
informations~\cite{Asul04,Shar04,Pan00,Lang02,McEl03,Vers04,Hana04,McEl05}.
However, it suffers from a lack of stability against temperature
and magnetic field changes, so a planar tunnel junction is a more
suitable device for temperature and magnetic field dependence studies.
Most ZBCP studies are on YBa$_2$Cu$_3$O$_{7-\delta}$ (Y-123) thin film
in planar tunnel junction experiments~\cite{Apri98,Apri99,Wang99,Shar02}.
Some of these films are highly oriented, thereby making the study on
angular dependence of ZBCPs possible~\cite{Iguc00}.
However, we chose to use Bi-2212 single crystals in tunneling
spectroscopic measurements, nevertheless, as the ZBCP is not commonly
available or studied on such mediums as Y-123.
Bi-2212 has several advantages in terms of properties
rather than other high-$T_\mathrm{c}$ cuprates.
For example, Bi-2212 is very stable against losing oxygen, retains
a cleavage property between BiO planes, possesses a large anisotropy
ratio between the $ab$-plane and $c$-axis directions, and its single
crystal does not twin like Y-123.

Bi-2212 single crystals were prepared by the traveling solvent
floating-zone (TSFZ) method, and the superconducting transition
temperature $T_\mathrm{c}$ of each as-grown single crystal was 
$87$--$90$ K, which was decided by the resistivity and magnetic
susceptibility measurements.
For hole doping concentration, as-grown single crystals are
located on slightly overdoped regions in the critical temperature-doping
phase diagram~\cite{Sato96}.
We have prepared planar tunnel junctions by using these single crystals.
Figure~\ref{fig:1}b shows a layout of a planar tunnel junction for the
$ab$-plane directions.
A single crystal with a size of approximately $10$ mm $\times$ $3$ mm
$\times$ $50$ $\mu$m was molded into a block of epoxy resin.
The block of epoxy resin was cut to expose the \{110\}-oriented surface
of the single crystals.
This cut surface was polished roughly by sandpaper, then smoothed by
diamond pastes with lubricant in the further process.
At that point, SiO was evaporated on the \{110\}-oriented surface
as a insulating barrier, and the thickness of SiO layer was modified
between $0$ {\AA} and $300$ {\AA} for each planar tunnel junction.
Ag was deposited on the SiO thin film through a metal mask patterned
with a counterelectrode of strips with width of $0.5$ mm.
The thickness of the Ag layer was controlled to be 1500 {\AA}.
For the planar type junctions made in this manner, Au wires were attached
to the Ag thin film with Ag paste, where the length $L$ between the
cross section of the tunnel junction and Ag paste regarded as the
reservoir in the circuit theory is roughly $1$--$2$ mm.
The total number of measured samples was more than 50.
In our experiments, the crystal orientation at the junction interface,
expressed as $\alpha$ in Figures~\ref{fig:1}a and \ref{fig:1}b, could
be predetermined from a sample rod configuration fabricated by the TSFZ
method since the sample rod grows up to the $a$-axis direction of
Bi-2212 single crystals in this method.
This fact was confirmed by taking into account the satellite reflections
arising from an incommensurate modulated structure in the X-ray
diffraction pattern of Bi-2212 single crystals~\cite{Idem90,Yama90}.

We have collected $I$--$V$ and $\mathrm{d}I/\mathrm{d}V$--$V$ data for the
prepared planar tunnel junctions by the standard $4$-terminal method.
The $\mathrm{d}I/\mathrm{d}V$--$V$ data corresponding to
$\sigma_\mathrm{T}(eV)$ were measured by using the conventional voltage
modulation method.
The temperature ranged from 4.2 K to 300 K, using a temperature
controller with a stability of more than 0.1 K.
Al-Al$_{2}$O$_{3}$-Pb planar tunnel junctions were fabricated and
their tunneling spectra were measured in advance, in order to compare
differences about the pairing symmetry of the superconducting order
parameter between conventional superconductors and high-$T_\mathrm{c}$
cuprate superconductors.

\section{Results and discussion}

\subsection{Tunneling conductance measurements}

\begin{figure}[tb]
\centerline{\includegraphics*[bb=50 50 510 815,width=8.3cm,keepaspectratio,clip]{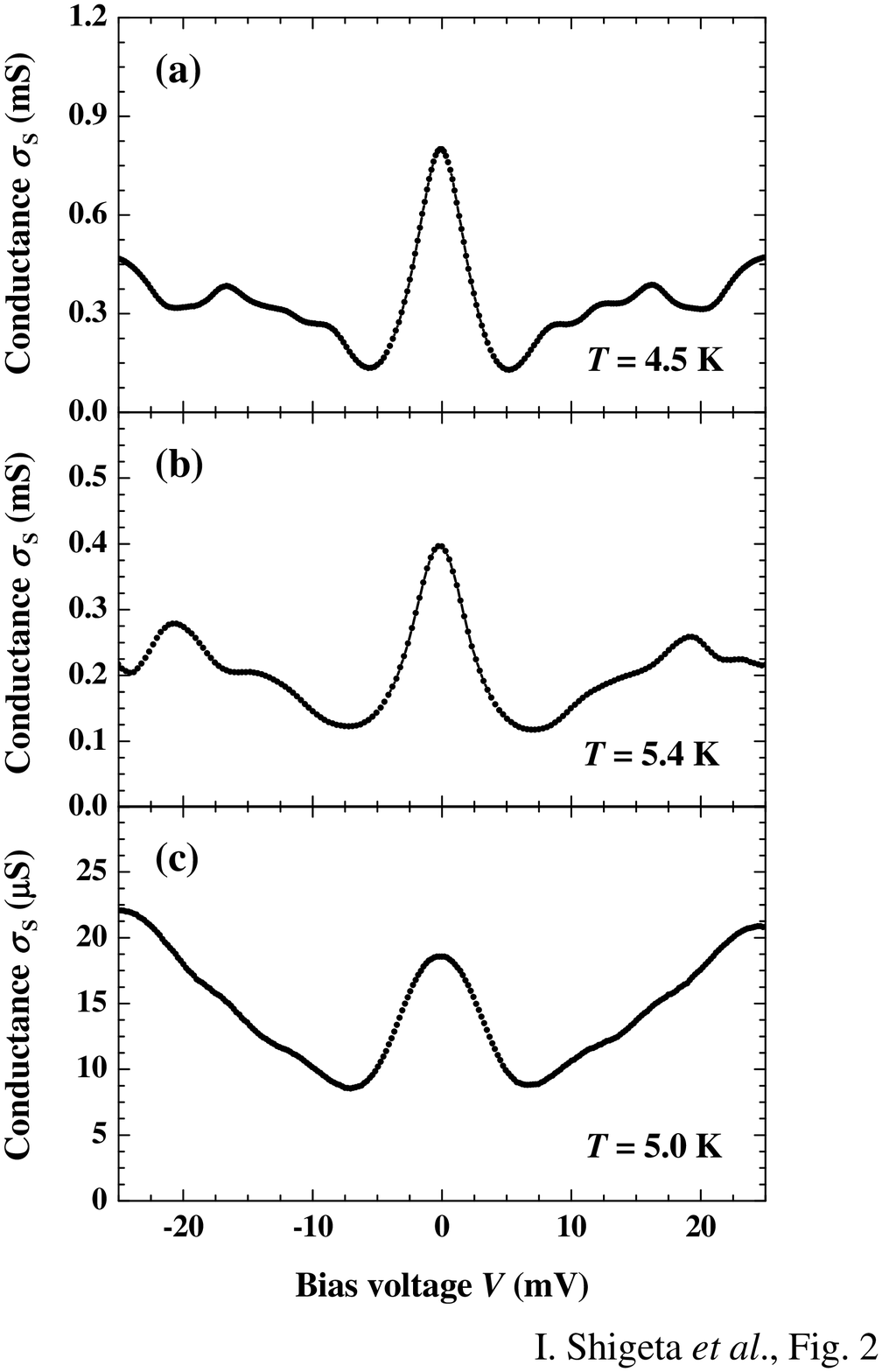}}
\caption{Comparison of typical spectral shapes for tunneling conductance
 $\sigma_\mathrm{S}(eV)$ in the several \{110\}-oriented tunnel
 junctions.
 An enhancement of the reproducible ZBCP was frequently observed below
 $T_\mathrm{c}$ in most of planar tunnel junctions for the
 \{110\}-oriented direction.
 There are the dip and weak modulation structures at the outside of the
 ZBCP.
\label{fig:2}}
\end{figure}

We have measured tunneling spectra of Ag-SiO-Bi-2212 planar tunnel
junctions.
Figure~\ref{fig:2} represents typical tunneling conductance
$\sigma_\mathrm{S}(eV)$ of the ZBCP in the superconducting state, which
was obtained on several \{110\}-oriented tunnel junctions.
The SiO tunnel barrier becomes gradually thicker from
Figure~\ref{fig:2}a to Figure~\ref{fig:2}c.
The values of tunneling conductance tend to enlarge with decreasing
thickness of the SiO barrier, but we cannot verify the definite
dependence of the ZBCP height on the insulating barrier thickness.
Then, there is no large difference between the \{110\}-oriented tunnel
junctions for concerning spectral shapes of the ZBCP, which was observed
in most of our planar tunnel junctions.
The ZBCP had a peak height of $2.0$--$3.3$ times higher than the
background and a full-width at half maximum (FWHM) of $3.3$--$6.5$ meV
in our experiments.
For the tunnel junctions that display the ZBCP, many peaks are higher
and sharper than the limit allowed by the BTK theory.
This ZBCP enhancement clearly indicates an intrinsic property of the
$d$-wave pairing symmetry in singlet superconductors.
As shown in Figure~\ref{fig:2}, both sides of the background around the
ZBCP gradually increase with leaving zero-bias voltage and have weak
modulation structures.
We conceive that the weak modulation structures appeared by the specific
current-path effect, due to the slight surface roughness at the junction
interface~\cite{Shig02}.
The coherent peaks at an energy gap region and any other specific
structures are not observed in tunneling conductance except for the ZBCP.

On the other hand, the typical $\sigma_\mathrm{T}(eV)$ tunneling into the
\{001\}-oriented surface, which was the cleavage surface of Bi-2212 single
crystals, had a well-known V-shaped gap structure~\cite{Shig00}.
The $d$-wave pairing symmetry of an order parameter is evidently
suggested from an anisotropy of the tunneling spectra, such as the
V-shaped gap structure in the \{001\}-oriented direction and the ZBCP in
the \{110\}-oriented direction.
In contrast to Ag-SiO-Bi-2212 planar tunnel junctions, a U-shaped gap
structure with an energy gap $\varDelta_{0} \simeq 1.4$ meV was observed
for Al-Al$_{2}$O$_{3}$-Pb planar tunnel junctions at $T = 1.8$ K,
where Pb was a superconducting state and Al was a normal state.
These tunnel junctions preserved resistances of more than several thousand
ohms, so we consider that said junctions correspond to ranges for high
potential barrier height as the Bardeen-Cooper-Schrieffer (BCS) limit in
the BTK theory, in which a tunneling spectral shape expects the U-shape
gap structure for $s$-wave superconductors.

\begin{figure}[tb]
\centerline{\includegraphics*[bb=35 160 535 640,width=8.3cm,keepaspectratio,clip]{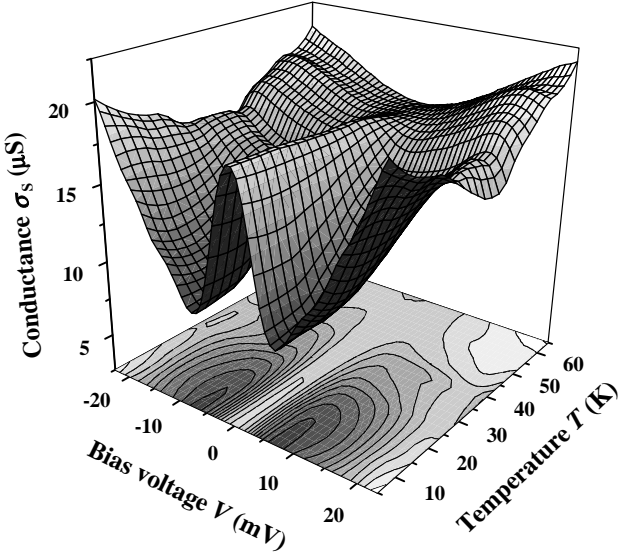}}
\caption{Temperature dependence of the ZBCP for the \{110\}-oriented
 tunnel junction in Figure~\ref{fig:2}c.
 As the temperature decreased, the ZBCP height increased, the ZBCP
 width sharpened, and the weak modulation structures enlarged.
 The ZBCP appeared below roughly $T = 60$ K. 
\label{fig:3}}
\end{figure}

Figure~\ref{fig:3} represents the temperature dependence of the ZBCP for
the \{110\}-oriented tunnel junction in Figure~\ref{fig:2}c.
As the temperature decreased, the ZBCP height enlarged, the ZBCP width
sharpened, the dip grew outside around the ZBCP, and the weak modulation
structures appeared around the ZBCP.
As shown in Figure~\ref{fig:3}, the ZBCP appeared below roughly 60 K.
In our tunneling experiments, the ZBCP appeared at $T_\mathrm{c}$ for
several tunnel junctions and below $T_\mathrm{c}$ for the other tunnel
junctions with the decrease in temperature.
Hence, the experimental results of the temperature dependence also denote
that the ZBCP is obviously caused by superconductivity.

Here, we will discuss other possible origins of the ZBCP in tunneling
conductance; for example, the magnetic impurity effect, known as the
Anderson-Appelbaum model taking account of the $s$-$d$ exchange
interaction~\cite{Appe66,Ande66,Appe67}, near the junction interface.
However, the possibility of the magnetic impurity effect is strongly
contradicted because the ZBCP enhancement has not been measured in the
\{001\}-oriented direction, of which the tunnel junctions were also
fabricated by using the same materials and procedures as those of the
\{110\}-oriented direction, as described in
Section~\ref{sec:experimental}.
Moreover, the ZBCP has been observed only below $T_\mathrm{c}$.
These facts obviously show that the ZBCP comes from not the magnetic
impurity effect, but the superconductivity of the $d$-wave pairing
symmetry.
Thus, the experimental results in Figures~\ref{fig:2} and \ref{fig:3}
coincide with the expectation of the MARS theory for $d$-wave
superconductors.
Furthermore, we will compare our experimental results with previous
research on planar tunnel junctions for Bi-2212 single
crystals~\cite{Deut05}.
S. Sinha \textit{et al.} reported that the ZBCP enhanced in $ab$-plane
directions at liquid-helium temperatures and disappeared above
$T_\mathrm{c}$ with increasing temperature~\cite{Sinh98a,Sinh98b}.
Their experimental results of the ZBCP behaviors are similar to ours
nevertheless the tunneling direction in the $ab$-plane at the junction
interface was not controlled in their experiments.

Another problem for the ZBCP remains, regarding the broken time-reversal
symmetry (BTRS) at a junction interface of anisotropic
superconductors~\cite{Covi97,Tana02,Kita03}.
The ZBCP must split below $T_\mathrm{c}$ if the BTRS occurs at the SIN
junction interface.
However, as shown in Figures~\ref{fig:2} and \ref{fig:3}, there is no
ZBCP splitting in all of the planar tunnel junctions for the
\{110\}-oriented  direction, on the condition of temperature ranges of
$4.2$--$300$ K not applied to magnetic fields.
Thus, we have deduced that the BTRS does not occur at the
high-$T_\mathrm{c}$ superconductor junction interface from our
experiments.

\subsection{Application to the circuit theory}

We will here discuss the smearing origin from a different point of view
against two theoretical analyses on the experimental ZBCP.
The actual ZBCP height depends on smearing processes of several possible
sources.
One of these substantial smearing effects is due to the increase
in the number of quasiparticles owing to thermal excitation.
Although the tunnel junctions were studied at relatively low
temperatures ($T/T_\mathrm{c} \simeq 0.06$, in the case of $T = 5.0$ K)
to minimize the effect of thermal excitations and to allow a more
straightforward deconvolution of the ZBCP structure, the ZBCP height was
not very high, when compared to the expectation of the ballistic theory
including the MARS mechanism~\cite{Tana95}.
In most of previous research, the origin of smearing processes was
usually explained as a result of the finite lifetime of quasiparticles
in superconductors by introducing the broadening factor $\varGamma$,
where $E$ is replaced by  $E - i\varGamma$ using the Dynes's method in
formulas of tunneling spectra~\cite{Dyne78,Dyne84}.
Here, the Dynes's parameter $\varGamma$ is introduced phenomenologically;
accordingly this method has not made clear how and where the finite
lifetime of quasiparticles comes from specifically.
However, this could be achieved to elucidate tunneling spectral
behaviors, sometimes even under changing temperatures or magnetic
fields, then used to fit theoretical curves of tunneling spectral
formulas with many experimental tunneling spectra up to the
present~\cite{Suzu94,Tana96,Alff97,Kash98,Cren00,Gonn01,Imai02,Schm02,Kohe03,Hoog03,Taka04}.

\begin{figure}[t]
\centerline{\includegraphics*[bb=30 210 560 641,width=8.3cm,keepaspectratio,clip]{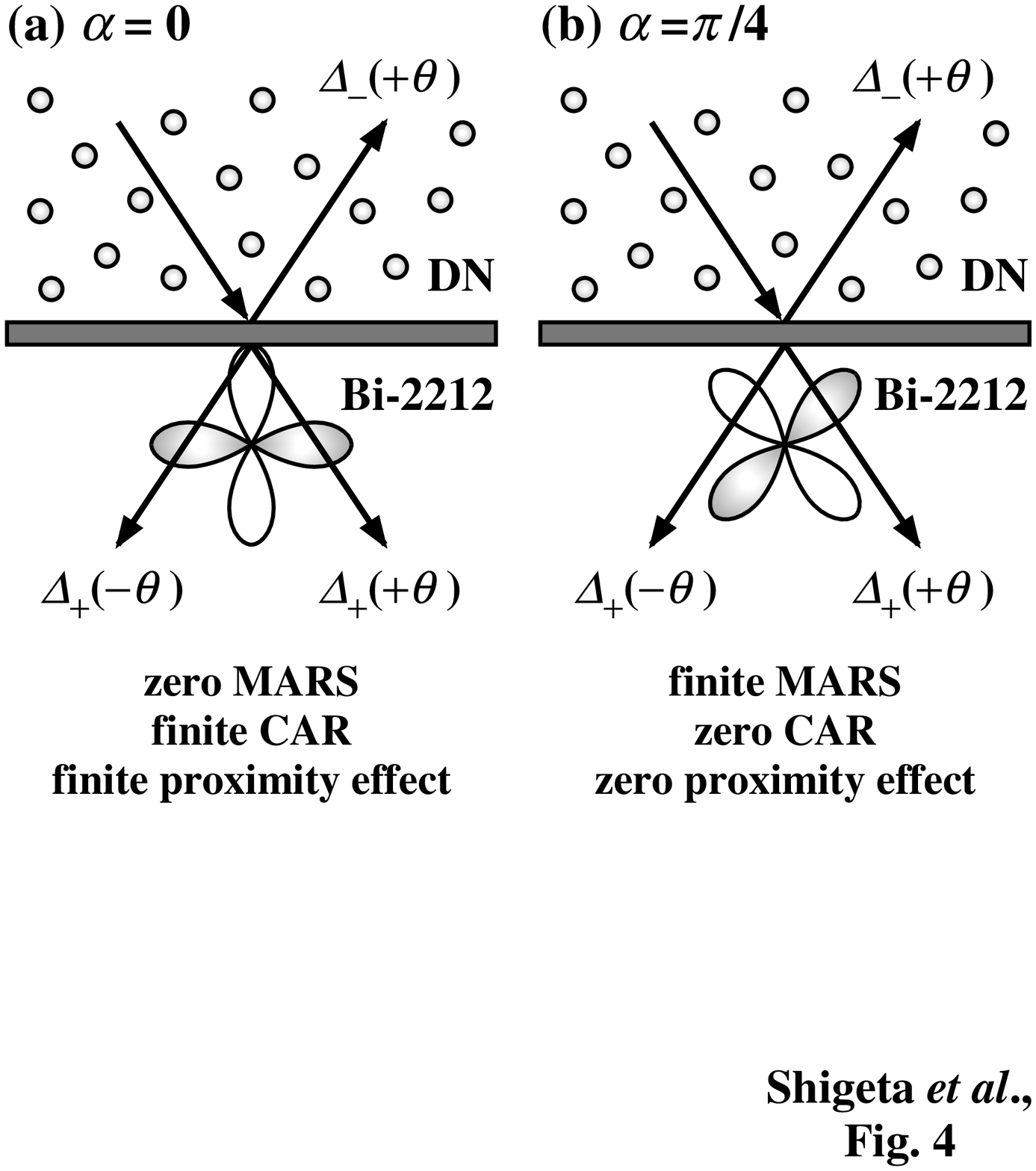}}
\caption{Schematic illustrations of the junction interface: 
 (a) $\alpha = 0$ and (b) $\alpha = \pi/4$.
 The trajectories in the scattering process for incoming and outgoing
 quasiparticles at the DN/I/S junction interface for $d$-wave
 superconductors and the corresponding pair potential 
 $\varDelta_{\pm}(\theta)$ are represented.
 We choose $\varDelta_{\pm} = \varDelta_{0}\cos[2(\theta \mp \alpha)]$
 for the pairing symmetry of an order parameter because of the 
 $d_{x^{2} - y^{2}}$-wave symmetry for Bi-2212 compounds.
 The our experimental condition corresponds to the case (b) for the
 situation at the junction interface.
\label{fig:4}}
\end{figure}
Up to this point, our experimental results for Ag-SiO-Bi-2212 planar
tunnel junctions reveal that Bi-2212 is evidently a $d$-wave
superconductor and that the ZBCP for \{110\}-oriented tunnel junctions
obviously arises from superconductivity.
We adapt the $d_{x^{2}-y^{2}}$-wave pairing symmetry for an order
parameter of Cooper pairs in Bi-2212 compounds for the analysis of the
circuit theory, because of the experimental evidence for many other
results in several measurement
methods~\cite{Ding96,Itoh94,Bour96,Pana98,Woll93}, as well as for the ZBCP
enhancement of the \{110\}-oriented direction in our tunneling
spectroscopic measurements.
Here, we choose $\alpha = \pi/4$ as the orientation of Bi-2212 single
crystals at the junction interface,
since the \{110\}-oriented surfaces were exposed at each junction
interface as shown in Figures~\ref{fig:2} and \ref{fig:3}.
For $\alpha = \pi/4$ the CAR and the proximity effect are completely
suppressed and only the MARS remain, while for $\alpha = 0$ the CAR and
the proximity effect are induced and the MARS is absent.
The situations at the DN/I/S junction interface are illustrated in
Figure~\ref{fig:4}.
Here, it is important to note as follows:
(i) It is valid for the restriction of only $\alpha = \pi/4$ in analysis
for the circuit theory because the MARS channels quench the CAR and the
proximity effect very effectively at  $\alpha > 0.02\pi$ for the
$d_{x^{2} - y^{2}}$-wave superconductor~\cite{Tana03a}.
(ii) The CAR and the proximity effect cannot influence against junction
properties in our fabricated junctions.
The mesoscopic interference effect, such as the CAR and the proximity
effect, must be negligible since our junction configuration satisfies $L
\gg L_\phi$, where the length  $L \simeq 1$--$2$ mm mentioned in
Section~\ref{sec:experimental} and the phase-breaking length 
$L_\phi \simeq 1$--$2$ $\mu$m in Ag thin films at liquid-helium
temperatures~\cite{Petr93,Petr95}.
In this case, quasiparticles cannot interfere each other in the whole of
the DN conductor on account of breaking the phase coherency between
quasiparticles before phase coherent quasiparticles reach the Ag paste
regarded as the reservoir~\cite{Wees92}.

As a result, the MARS channels sufficiently give influences in line
shapes of $\sigma_\mathrm{T}(eV)$ rather than the CAR and the proximity
effect in wide directions around  $\alpha = \pi/4$,
even if actual tunnel junctions slightly retain the surface roughness at
the junction interface.
Therefore, in analysis for around $\alpha = \pi/4$ corresponding to 
$\alpha \gg 0.02\pi$, those facts justify usage of a simplified formula
for the circuit theory, just expressed by the Ohm's law: an elementary
sum of $R_{R_\mathrm{d}=0}$ and $R_\mathrm{d}$ in a superconducting
state.
Without solving the Usadel equation in such a case, the normalized
tunneling conductance $\sigma_\mathrm{T}(eV)$ can be approximated by the
following simplified equations~\cite{Tana04a}:
\begin{gather}
\sigma_\mathrm{T}(eV) 
  = \frac{\sigma_\mathrm{S}(eV)}{\sigma_\mathrm{N}(eV)}
  = \frac{R_\mathrm{b} + R_\mathrm{d}}
         {R_{R_\mathrm{d}=0} + R_\mathrm{d}}, \; 
\label{eq:1} \\
R_{R_\mathrm{d}=0} = \frac{R_\mathrm{b}}{\langle I_\mathrm{b0}
 \rangle}, \;
\langle I_\mathrm{b0} \rangle 
  = \frac{\langle I_\mathrm{b0} \rangle_\mathrm{S}}
         {\langle I_\mathrm{b0} \rangle_\mathrm{N}}, 
\label{eq:2} \\
\langle I_\mathrm{b0} \rangle_\mathrm{S}
  = \frac{1}{4k_\mathrm{B}T}
    \int^{\infty}_{-\infty}\mathrm{d}E
    \frac{1}{\pi}
    \int^{\pi/2}_{-\pi/2}\mathrm{d}\theta \; e^{-\lambda\theta^{2}} 
    \hspace*{8ex} \nonumber \\
    \hspace*{16ex} \times \;
    I_\mathrm{b0} \cos\theta \; \mathrm{sech}^{2} 
    \left( \frac{E +  eV}{2k_\mathrm{B}T} \right), 
\label{eq:3} \\
\langle I_\mathrm{b0} \rangle_\mathrm{N}
  = \frac{1}{4k_\mathrm{B}T}
    \int^{\infty}_{-\infty}\mathrm{d}E
    \frac{1}{\pi}
    \int^{\pi/2}_{-\pi/2}\mathrm{d}\theta \; e^{-\lambda\theta^{2}} 
    \hspace*{8ex} \nonumber \\
    \hspace*{16ex} \times \;
    T(\theta) \cos\theta \; \mathrm{sech}^{2} 
    \left( \frac{E +  eV}{2k_\mathrm{B}T} \right), 
\label{eq:4} \\
I_\mathrm{b0} 
  = \frac{ T(\theta) \left\{ 1 + T(\theta) |\varGamma_{+}|^{2} 
         + \left[ T(\theta) - 1 \right] |\varGamma_{+}\varGamma_{-}|^{2} 
         \right\}}
         {|1 + \left[ T(\theta) - 1 \right]
         \varGamma_{+}\varGamma_{-}|^{2}}, 
\label{eq:5} \\
T(\theta) = \frac{4\cos^{2}\theta}{4\cos^{2}\theta + Z^{2}}, 
\label{eq:6} \\
\varGamma_{+} = \frac{\varDelta^{*}_{+}}
                     {E + \sqrt{E^{2} - \varDelta^{2}_{+}}}, \;
\varGamma_{-} = \frac{\varDelta_{-}}
                     {E + \sqrt{E^{2} - \varDelta^{2}_{-}}},
\label{eq:7}
\end{gather}
with $\varDelta_{\pm} = \varDelta_{0}\cos[2(\theta \mp \alpha)]$, where
$\varDelta_{0}$ denotes the maximum amplitude of the pair potential and
$\theta$ is the injection angle of the quasiparticle measured from the
$x$-axis.
Here, an insulating barrier is expressed as a $\delta$-function model
$H\delta(x)$, where $Z$ is an insulating barrier height given by $Z =
2mH/(\hbar^{2}k_\mathrm{F})$ with Fermi momentum $k_\mathrm{F}$ and
effective mass $m$.
In the above equations, $k_\mathrm{B}$ is the Boltzmann's constant and
the factor $\cos\theta$ is necessary for calculating a normal component
relative to the junction interface of the tunneling current.
Further $\lambda$ is added in order to introduce the probability of
tunneling directions at the junction interface, where we presume to be
the Gauss distribution as a weight function.

\begin{figure}[t]
\centerline{\includegraphics*[bb=40 225 540 630,width=8.3cm,keepaspectratio,clip]{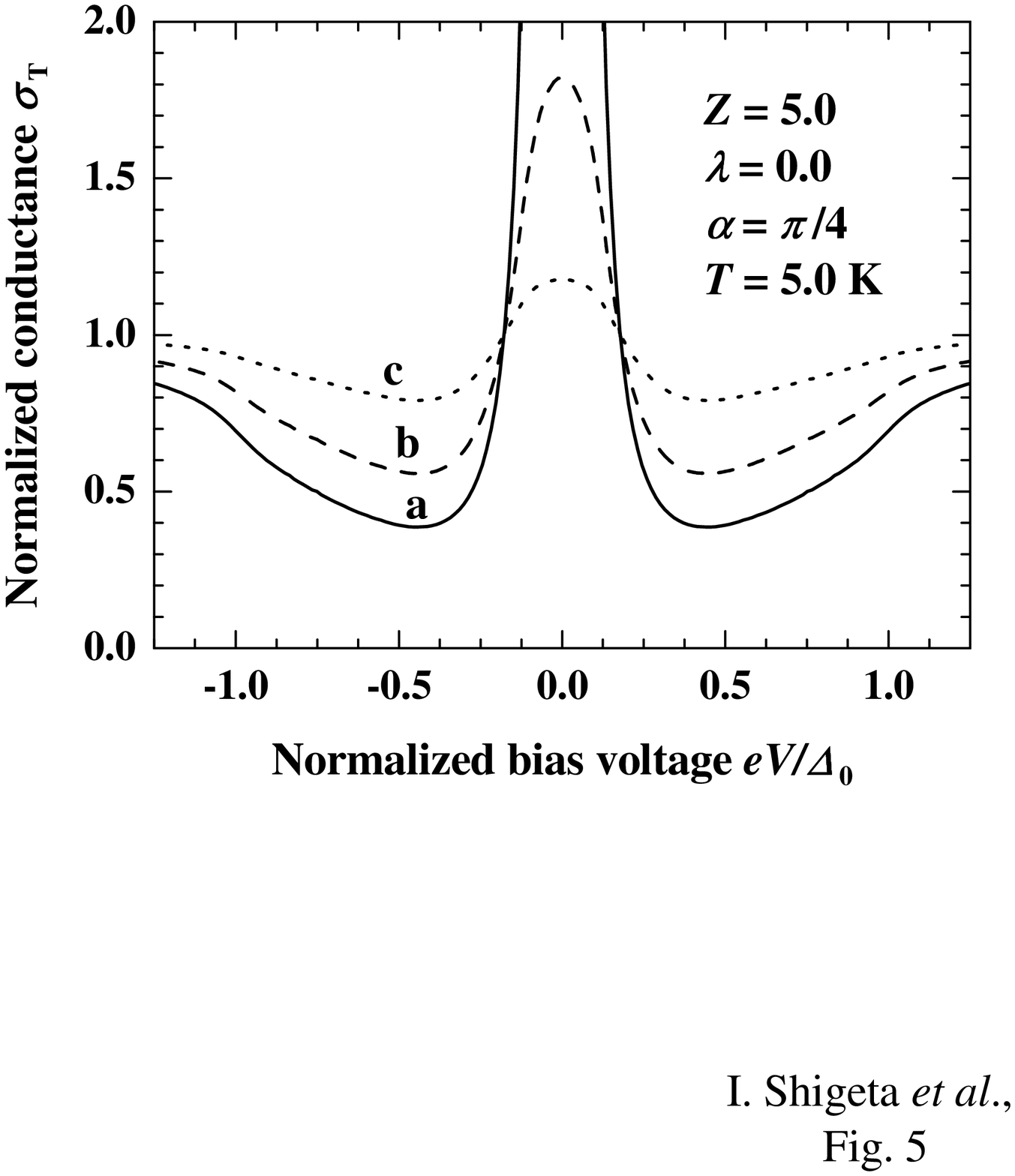}}
\caption{
 Variations of the normalized tunneling conductance
 $\sigma_\mathrm{T}(eV)$ calculated by the circuit theory under the
 condition of $Z = 5.0$, $\lambda = 0.0$ and $\alpha = \pi/4$ at $T =
 5.0$ K.
 a: $R_\mathrm{d}/R_\mathrm{b} = 0.0$, b: $R_\mathrm{d}/R_\mathrm{b} =
 1.0$ and c: $R_\mathrm{d}/R_\mathrm{b} = 5.0$.
 It is important that the DN resistance $R_\mathrm{d}$
 effectively suppresses the ZBCP height but does not change the ZBCP
 width on the condition of $\alpha = \pi/4$.
 This is because the ZBCP comes from the formation of the MARS.
\label{fig:5}}
\end{figure}

\begin{figure}[t]
\centerline{\includegraphics*[bb=40 225 540 630,width=8.3cm,keepaspectratio,clip]{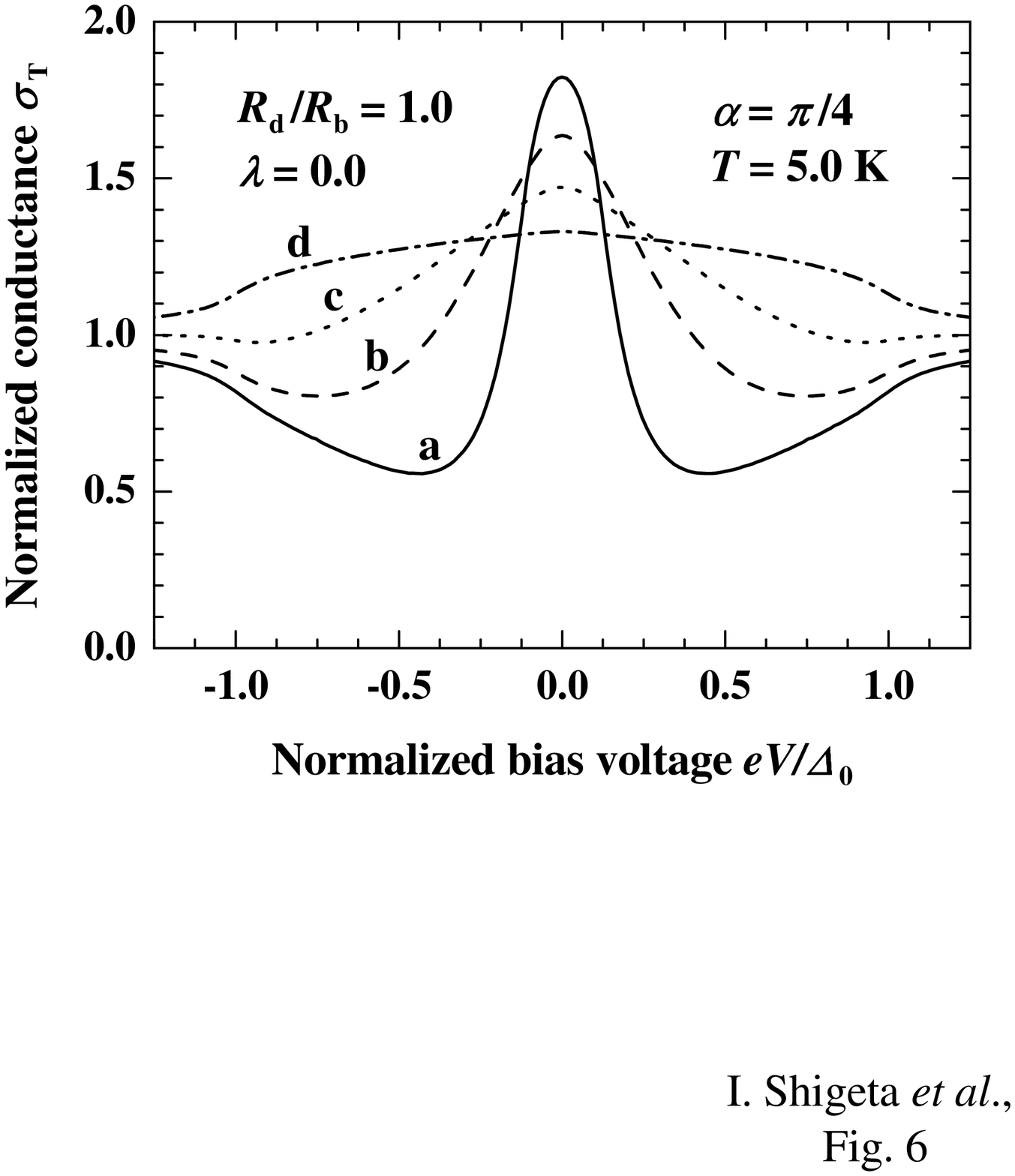}}
\caption{
 Variations of the normalized tunneling conductance
 $\sigma_\mathrm{T}(eV)$ calculated by the circuit theory under the
 condition of $R_\mathrm{d}/R_\mathrm{b}= 1.0$, $\lambda = 0.0$ and 
 $\alpha = \pi/4$ at $T = 5.0$ K.
 a: $Z = 5.0$, b: $Z = 2.0$, c: $Z = 1.0$ and d: $Z = 0.0$.
\label{fig:6}}
\end{figure}

\begin{figure}[t]
\centerline{\includegraphics*[bb=40 225 540 630,width=8.3cm,keepaspectratio,clip]{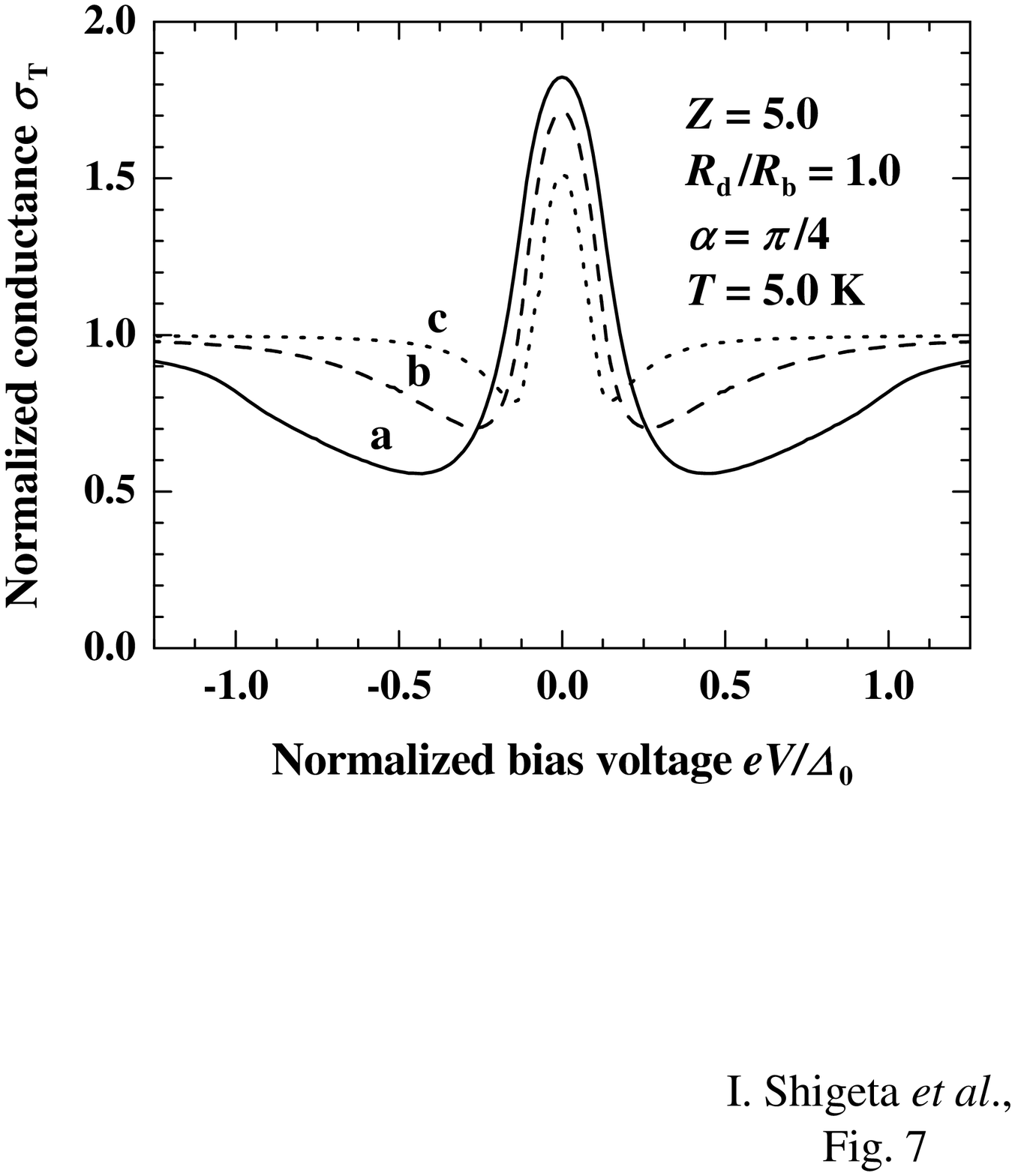}}
\caption{
 Variations of the normalized tunneling conductance
 $\sigma_\mathrm{T}(eV)$ calculated by the circuit theory under the
 condition of $Z = 5.0$, $R_\mathrm{d}/R_\mathrm{b}= 1.0$ and 
 $\alpha = \pi/4$ at $T = 5.0$ K.
 a: $\lambda = 0.0$, b: $\lambda = 10.0$ and c: $\lambda = 90.0$.
\label{fig:7}}
\end{figure}

The recent theoretical expectations of the circuit theory motivate us to
analyze experimental ZBCP spectra in the $ab$-plane directions.
For the circuit theory with the $d$-wave pairing symmetry, the
theoretical results indicate that the ZBCP height is strongly reduced by
taking into account the existence of the DN conductor, and the resulting
$\sigma_\mathrm{T}(0)$ is not as high as that obtained in the ballistic
regime~\cite{Tana03a,Tana04a}.
One of the typical examples is plotted in Figure~\ref{fig:5}.
In the case of $\alpha = \pi/4$, the line shapes of
$\sigma_\mathrm{T}(eV)$ are independent of $E_\mathrm{Th}$ because of
the absence of the CAR and the proximity effect.
The parameter values change only for $R_\mathrm{d}/R_\mathrm{b}$ in the
tunneling spectral formula of equations~(\ref{eq:1}--\ref{eq:7}).
As shown in Figure~\ref{fig:5}, the effect of $R_\mathrm{d}$ is obviously
significant for the ZBCP height of the resulting
$\sigma_\mathrm{T}(eV)$, but is independent of the ZBCP width of the
resulting $\sigma_\mathrm{T}(eV)$.
On the other hand, the ZBCP width is determined by an amplitude of
$\varDelta_{0}$, $Z$ and $\lambda$ on the condition of $\alpha =
\pi/4$.
Figures~\ref{fig:6} and \ref{fig:7} represent typical variations of line
shapes of $\sigma_\mathrm{T}(eV)$ for $Z$ and $\lambda$, respectively.
Different from the case of $R_\mathrm{d}$ in Figure~\ref{fig:5}, the
height and width of the ZBCP change simultaneously with alterations to the
value of $Z$ or $\lambda$.
Here we note that, for the circuit theory, $R_\mathrm{b}$ depends on
$Z$ by way of equation~(\ref{eq:6}) and also the following
equation~\cite{Tana04a}: 
\begin{equation}
 R_\mathrm{b} 
  = \frac{2R_{0}}{\displaystyle \int^{\pi/2}_{-\pi/2}
    \mathrm{d}\theta \; T(\theta) \cos\theta},
\label{eq:8}
\end{equation}
which has Sharvin resistance $R_{0}$ at the junction interface.
Therefore, for the actual quantitative comparison with tunneling
experiments, these facts suggest that we must take into account the
effect of $R_\mathrm{d}$.
As shown in Figure~\ref{fig:2}, the resulting ZBCP height from experiments
is not as high as that obtained in the ballistic regime, so we can
expect that the influence of $R_\mathrm{d}$ can replace that of Dynes's
parameter $\varGamma$~\cite{Dyne78,Dyne84} as a broadening factor in
tunneling spectral formulas for the ZBCP.
Therefore, we will analyze our experimental ZBCP results in terms of the
circuit theory extended to the systems containing $d$-wave
superconductor tunnel junctions.

\subsection{Fitting results for the circuit theory}
\begin{figure}[t]
\centerline{\includegraphics*[bb=40 225 540 630,width=8.3cm,keepaspectratio,clip]{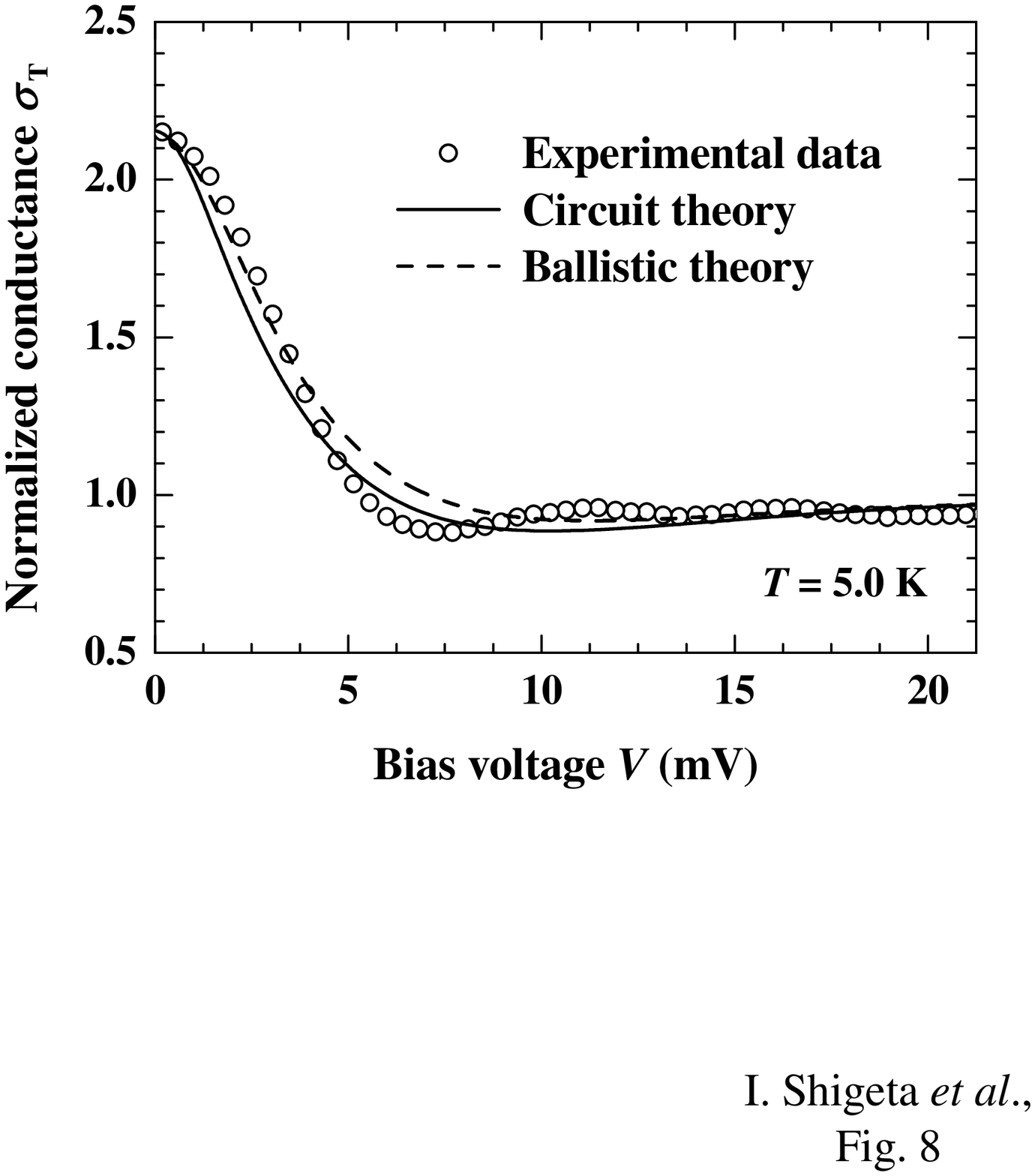}}
\caption{
 Experimental tunneling spectrum at $T = 5.0$ K with the
 tunneling spectra of the circuit and ballistic theories for $d$-wave
 superconductors.
 The open circles represent the experimental result of the
 \{110\}-oriented junction for the Bi-2212 single crystal in
 Figure~\ref{fig:2}c.
 The experimental spectrum is normalized by a parabolic background.
 The solid line represents the tunneling spectrum calculated by the
 circuit theory extended to the $d$-wave pairing symmetry; and the broken
 line, that of the ballistic theory extended as well.
\label{fig:8}}
\end{figure}

Figure~\ref{fig:8} shows the spectral fit of the circuit theory for
$d$-wave superconductors to our experimental data.
In addition, in order to make a comparison between the circuit theory
for DN/I/S tunnel junctions and the ballistic theory for N/I/S tunnel
junctions, the fitting result for the ballistic theory is also given
in Figure~\ref{fig:8}.
This result was found by calculating the equations cited in
reference~\cite{Shig00}.
The open circles represent the experimental data for the \{110\}-oriented 
tunnel junction in Figure~\ref{fig:2}c, normalized by a parabolic
background.
The solid line represents the fitting curve for the circuit theory
extended to the systems containing $d$-wave superconductor tunnel junctions.
The broken line represents that for the ballistic theory extended for
$d$-wave superconductors, as well.
The fitting results indicate good agreement between the experimental
data and two theoretical curves.
Other tunnel junctions of the same compound have been also studied in
this fashion, and the results are very reproducible.
The slight discrepancy between the experimental data and theoretical
calculations in Figure~\ref{fig:8} seems to come from the simplification
of the insulating barrier in the theories.
Here, a junction interface between a superconducting state and a normal
state is clearly defined, and a $\delta$-function is used as a potential
barrier.
An actual tunnel junction is more complicated.

Table~\ref{tab:table1} denotes the corresponding parameter values in
each theory.
The difference in two fitting results is essentially recapitulated in
values of $\varGamma$ and $R_\mathrm{d}/R_\mathrm{b}$.
Here, we emphasize that an introduction of the Dynes's parameter
$\varGamma$ in tunneling spectral formulas may not necessarily suggest
the concrete smearing origin in experimental tunneling spectra in the
case of (B) for the ballistic theory.
On the other hand, in the analysis in the case of (A) for the circuit
theory, the Dynes's parameter $\varGamma$ as the broadening factor does
not need to be introduced into the tunneling spectral formula because
the resistance $R_\mathrm{d}$ of the DN conductor plays an important
role as the suppression of the ZBCP height in tunneling conductance.
\begin{table}[t]
\caption{Typical values of the fitting parameters in each tunneling
 conductance formula of (A) the circuit theory and (B) the ballistic
 theory, when compared to the experimental result for the
 \{110\}-oriented tunnel junction in Figure~\ref{fig:2}c.
 The ``---'' means that the parameter as $R_\mathrm{d}/R_\mathrm{b}$
 does not exist in the spectral formula of the ballistic theory.
 It is important to note that, for the circuit theory, the $\varGamma$
 value as the broadening factor can equal to 0.0 meV, in spite of
 obtaining good fitting results.
\label{tab:table1}}
\begin{tabular}{ccccccc}
\hline\noalign{\smallskip}  
& $\varDelta_{0}$ (meV) & $\varGamma$ (meV) & $Z$ & 
$\lambda$ & $R_\mathrm{d}/R_\mathrm{b}$ & $\alpha$ \\
\noalign{\smallskip}\hline\noalign{\smallskip}  
(A) & 25.0 & 0.0 & 2.2 & 13.0 & 0.35 & $\pi$/4 \\
(B) & 22.0 & 1.7 & 2.3 & 13.0 & ---  & $\pi$/4 \\
\noalign{\smallskip}\hline  
\end{tabular}
\end{table}

\begin{figure}[t]
\centerline{\includegraphics*[bb=40 225 540 630,width=8.3cm,keepaspectratio,clip]{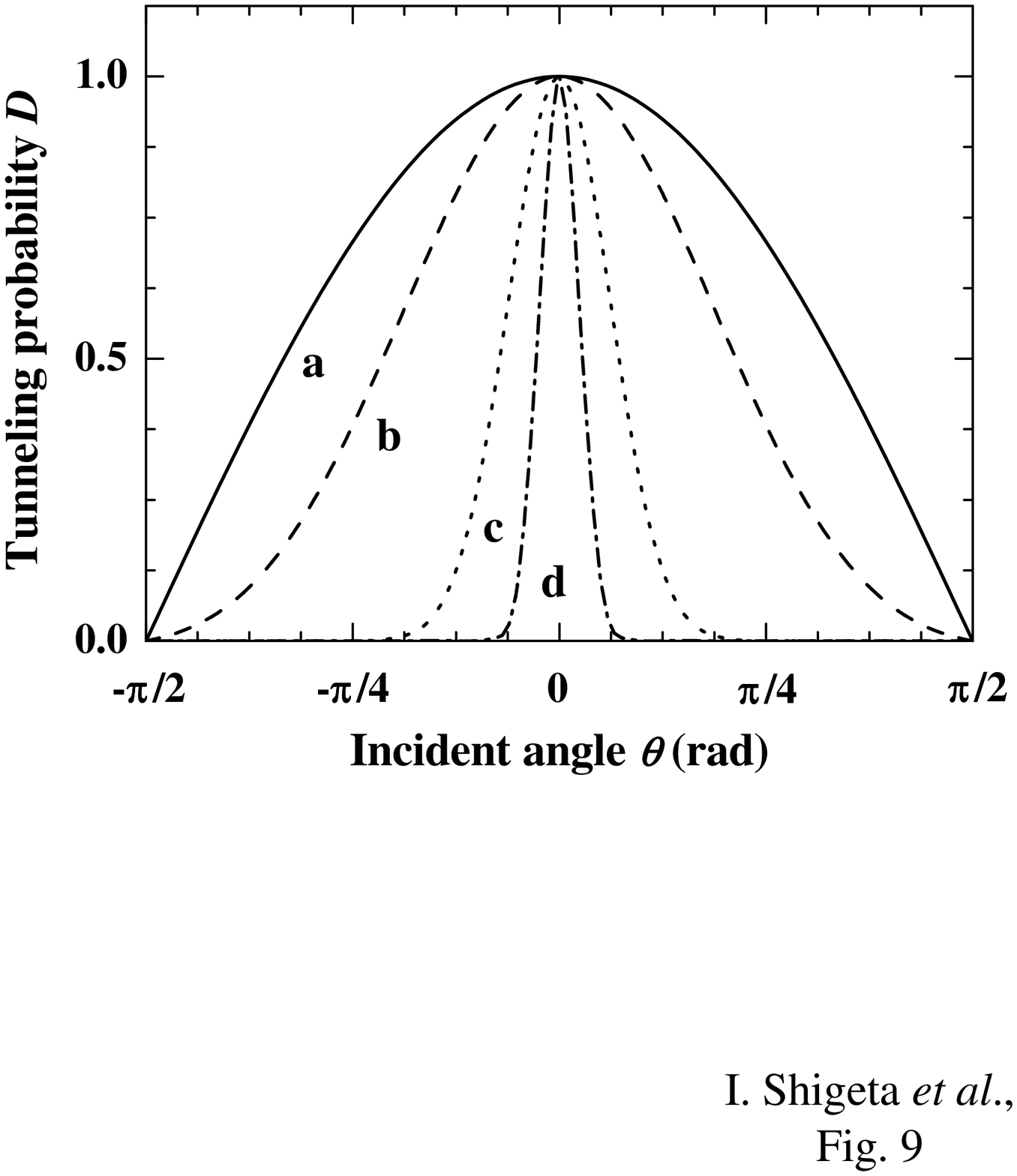}}
\caption{Tunneling probability $D$ as a function of incident angle
 $\theta$ for a: $\lambda = 0.0$, b: $\lambda = 1.0$, c: $\lambda =
 13.0$ and d: $\lambda = 90.0$.
 The dotted line c is corresponding to the fitting results in both of the
 circuit theory and the ballistic theory in Figure~\ref{fig:8}.
\label{fig:9}}
\end{figure}

Here, we note several features for the fitting parameter values obtained
from the circuit theory.
The amplitude of $\varDelta_{0} = 25.0$ meV in the \{110\}-oriented
direction is appropriate for comparison to ranges of roughly 
$\varDelta_{0} = 16$--$40$ meV for energy gap values in the
\{001\}-oriented direction~\cite{Huan89,Miya99} and to the value of
$\varDelta_{0} = 22.0$ meV obtained through the analysis of the
\{110\}-oriented direction by using the ballistic theory in
Table~\ref{tab:table1}.
Thus, for $2\varDelta_{0}/k_\mathrm{B}T_\mathrm{c}$ a value of about 6.5
was obtained, typical for high-temperature superconductors~\cite{Kita96}.
The value of $Z = 2.2$ is comparatively small in variable ranges
(see Fig.~10 in Ref.~\cite{Tana04a}), so this result corresponds
to relatively the low barrier height for our tunnel junctions.
Corresponding to a finite value of $Z$, a finite potential height exists
at the junction interface.
Therefore, the tunneling directions of electrons and holes also restrict
around normal to the junction interface.
Consequently, the finite value of $\lambda = 13.0$ was obtained.
Furthermore, we consider how the $\sigma_\mathrm{T}(eV)$ depends on the
spread of $\vec{k}$-space in which tunneling electrons reach.
The tunneling probability $D$ as a function of the incident angle
$\theta$ results in $D \propto \exp(-\lambda \theta^{2}) \cos\theta$
when $U \gg E_\mathrm{F}$, where $U$ is the height of a tunneling
barrier and $E_\mathrm{F}$ is Fermi energy~\cite{Suzu01}.
The term $\exp(-\lambda \theta^{2}) \cos\theta$ is also contained in the
equations cited in references~\cite{Shig00,Kash95}.
As seen in Figure~\ref{fig:9}, the directional dependence of tunneling
probability becomes strong with increasing $\lambda$.
When $\lambda = 0$, tunneling electrons range most in $\vec{k}$-space.
On the other hand, when $\lambda$ is enough large, electrons are
restricted in the narrow region along an $x$-axis through the $\varGamma$
point in the Brillouin zone.
The planar type junction detects the electric state in narrow
$\vec{k}$-space, while the point contact spectroscopy and the
STM/STS detect the electric state in wide $\vec{k}$-space.
Therefore, it is reasonable that we have estimated $\lambda = 13.0$ as
the parameter of the $D$ from the analysis of the circuit theory for
$d$-wave superconductors.
Furthermore, the ratio of $R_\mathrm{d}/R_\mathrm{b} = 0.35$ may be not
so large, but this value is sufficient to smear the spectral shape of
the ZBCP by the influence of $R_\mathrm{d}$ in systems of $d$-wave
superconductor tunnel junctions because the theoretical ZBCP height is
effectively suppressed as shown in Figure~\ref{fig:5}.
As a result, in Figure~\ref{fig:8}, we received fitting results
between the experimental ZBCP result in the \{110\}-oriented directions
and the circuit theory.
Hence, through analysis of the circuit theory, we are able not only
to exclude the Dynes's parameter $\varGamma$ from the tunneling spectral
formula, but also able to obtain good fitting results of the ZBCP for
the \{110\}-oriented tunnel junctions.

In this situation, we were successful in explaining the ZBCP
behaviors by using the circuit theory for $\alpha = \pi/4$ including
only the MARS and the influence of $R_\mathrm{d}$ in the tunneling
spectral formula, in which the CAR and the proximity effect are
completely suppressed.
Other mechanisms may be responsible for the observed ZBCP smearing but
the theoretical and quantitative estimates are certainly consistent with
what we know about the DN resistance $R_\mathrm{d}$.
We emphasize that it is appropriate for tunneling spectral formula to
introduce the combined resistance for 
$R_{R_\mathrm{d}=0} + R_\mathrm{d}$, rather than the artificial
smearing of $\varGamma$.
It is suitable to understand that there is the relation between
$R_\mathrm{d}$ and $\varGamma$.
Namely, we can comprehend that finite lifetime of quasiparticles due to
Dynes's parameter $\varGamma$ arises from the scattering of
quasiparticles at scatters in the DN conductor, which is expressed by
$R_\mathrm{d}$ in equations~(\ref{eq:1}--\ref{eq:7}) of the circuit theory.
Thus, we can conclude that the effect of $R_\mathrm{d}$ plays an
important role in tunneling spectra of high-$T_\mathrm{c}$ cuprate
tunnel junctions as a new point of view for smearing origins in spectral
shapes of the ZBCP.
However, in order to progress further with analysis for the tunneling
spectra in anisotropic superconductors, one must solve the Usadel
equations, such as in reference~\cite{Tana04a}, for the quantitative
discussions including much more general cases with arbitrary $\alpha$
values.

\section{Conclusions}

We have measured Ag-SiO-Bi-2212 planar tunnel junctions.
The ZBCP enhancement was observed below $T_\mathrm{c}$ for the
\{110\}-oriented tunnel junctions.
Our experimental results for Ag-SiO-Bi-2212 planar tunnel junctions
denote that Bi-2212 is obviously a $d$-wave superconductor and that the
ZBCP originates from superconductivity.
Hence, we have analyzed the ZBCP behaviors by using the circuit theory,
which is constructed under the condition of the elementary sum of 
$R_{R_\mathrm{d}=0}$ and $R_\mathrm{d}$ at the DN/I/S junction
interface for $d$-wave superconductors.
The fitting results indicated good agreements between the experimental 
data and the theoretical curves not taking into account the Dynes's
broadening factor $\varGamma$, in spite of the extreme case of 
$\alpha = \pi/4$ including only the influence of the MARS and
$R_\mathrm{d}$ in the formula of tunneling conductance.
As a new interpretation about smearing origins, the circuit theory gives
that  the smearing factor in tunneling spectra could be understood as the
effect of the DN conductor in tunnel junctions, rather than that of
Dynes's broadening parameter $\varGamma$.
Thus, the circuit theory well describes the phenomena of quasiparticle
tunnelings in anisotropic superconductors, rather than the ballistic
theory, and such is very applicable to analysis of the actual
experimental data of tunneling spectra of high-$T_\mathrm{c}$ cuprate
tunnel junctions.

\section*{Acknowledgments}
\label{sec:acknowledgments}
We would like to thank T. Arai, T. Uchida and Y. Tominari for
supplying Bi-2212 single crystals.
We appreciate T. Aomine, T. Fukami, S. W. M. Scott, S. Keenan, M. Briney
and A. Odahara for helpful advice and discussions.
We are additionally grateful to T. Asano for supporting SQUID
measurements on some single crystals.

%

\begin{thebibliography}{99}
%
%

\bibitem{Lesu92} J. Lesueur, L. H. Greene, W. L. Feldmann, A. Inam, 
        Physica C \textbf{191}, 325 (1992)
\bibitem{Tair98} M. Taira, M. Suzuki, X.-G. Zheng, T. Hoshino, 
	J. Phys. Soc. Jpn. \textbf{67}, 1732 (1998)
\bibitem{Wei98} J. Y. T. Wei, N.-C. Yeh, D. F. Grarrigus, M. Strasik, 
        Phys. Rev. Lett. \textbf{81}, 2542 (1998)
\bibitem{Kash00} S. Kashiwaya, Y. Tanaka, 
        Rep. Prog. Phys. \textbf{63}, 1641 (2000), and references
        therein
\bibitem{Suzu01} A. Suzuki, M. Taira, M. Suzuki and X.-G. Zheng,  
        J. Phys. Soc. Jpn. \textbf{70}, 3018 (2001)
\bibitem{Mao01} Z. Q. Mao, K. D. Nelson, R. Jin, Y. Liu, Y. Maeno, 
        Phys. Rev. Lett. \textbf{87}, 037003 (2001)
\bibitem{Shar01} A. Sharoni, G. Koren, O. Millo, 
        Europhys. Lett. \textbf{54}, 675 (2001)
\bibitem{Lofw01} T. L\"{o}fwander, V. S. Shumeiko, G. Wendin,
        Supercond. Sci. Technol. \textbf{14}, R53 (2001), and references
        therein
\bibitem{Shig02} I. Shigeta, F. Ichikawa, T. Aomine, 
        Physica C \textbf{378-381}, 316 (2002)
\bibitem{Frea03a} M. Freamat, K.-W. Ng, 
        Physica C \textbf{400}, 1 (2003)
\bibitem{Mao03} Z. Q. Mao, M. M. Rosario, K. D. Nelson, K. Wu, I. G. Deac, 
        P. Schiffer, Y. Liu, T. He, K. A. Regan, R. J. Cava, 
        Phys. Rev. B \textbf{67}, 094502 (2003)
\bibitem{Qazi03} M. M. Qazilbash, A. Biswas, Y. Dagan, R. A. Ott,
        R. L. Greene, 
        Phys. Rev. B \textbf{68}, 024502 (2003)
\bibitem{Frea03b} M. Freamat, K.-W. Ng, 
        Phys. Rev. B \textbf{68}, 060507 (2003)
\bibitem{Miya03} T. Miyake, T. Imaizumi, I. Iguchi, 
        Phys. Rev. B \textbf{68}, 214520 (2003)
\bibitem{Kash04} H. Kashiwaya, S. Kashiwaya, B. Prijamboedi, A. Sawa,
        I. Kurosawa, Y. Tanaka, I. Iguchi,
        Phys. Rev. B \textbf{70}, 094501 (2004)
\bibitem{Kawa05} M. Kawamura, H. Yaguchi, N. Kikugawa, Y. Maeno, 
        H. Takayanagi, 
	J. Phys. Soc. Jpn. \textbf{74}, 531 (2005)
\bibitem{Deut05} G. Deutscher,
        Rev. Mod. Phys. \textbf{77}, 109 (2005), and references therein
\bibitem{Appe66} J. Appelbaum, 
        Phys. Rev. Lett. \textbf{17}, 91 (1966)
\bibitem{Ande66} P. W. Anderson, 
        Phys. Rev. Lett. \textbf{17}, 95 (1966)
\bibitem{Appe67} J. A. Appelbaum, 
        Phys. Rev. \textbf{154}, 633 (1967)
\bibitem{Sand94} S. C. Sanders, S. E. Russek, C. C. Clickner, J. W. Ekin,
        Appl. Phys. Lett. \textbf{65}, 2232 (1994)
\bibitem{Covi96} M. Covington, R. Scheuerer, K. Bloom, L. H. Greene, 
        Appl. Phys. Lett. \textbf{68}, 1717 (1996)
\bibitem{Tana95} Y. Tanaka, S. Kashiwaya, 
        Phys. Rev. Lett. \textbf{74}, 3451 (1995)
\bibitem{Blon82} G. E. Blonder, M. Tinkham, T. M. Klapwijk, 
        Phys. Rev. B \textbf{25}, 4515 (1982)
\bibitem{Volk93} A. F. Volkov, A. V. Zaitsev, T. M. Klapwijk, 
        Physica C \textbf{210}, 21 (1993)
\bibitem{Naza94} Yu. V. Nazarov, 
        Phys. Rev. Lett. \textbf{73}, 1420 (1994)
\bibitem{Naza99} Yu. V. Nazarov, 
        Superlattices Microstruct. \textbf{25}, 1221 (1999)
\bibitem{Tana03a} Y. Tanaka, Yu. V. Nazarov, S. Kashiwaya, 
        Phys. Rev. Lett. \textbf{90}, 167003 (2003)
\bibitem{Tana03b} Y. Tanaka, A. A. Golubov, S. Kashiwaya, 
        Phys. Rev. B \textbf{68}, 054513 (2003)
\bibitem{Tana04a} Y. Tanaka, Yu. V. Nazarov, A. A. Golubov, 
        S. Kashiwaya, 
        Phys. Rev. B \textbf{69}, 144519 (2004); 
        Y. Tanaka, Yu. V. Nazarov, A. A. Golubov, S. Kashiwaya, 
        Phys. Rev. B \textbf{70}, 219907(E) (2004)
\bibitem{Tana04b} Y. Tanaka, S. Kashiwaya, 
        Phys. Rev. B \textbf{70}, 012507 (2004)
\bibitem{Yoko05a} T. Yokoyama, Y. Tanaka, A. A. Golubov, J. Inoue,
	Y. Asano,
        Phys. Rev. B \textbf{72}, 094506 (2005)
\bibitem{Tana05} Y. Tanaka, Y. Asano, A. A. Golubov, S. Kashiwaya,
        Phys. Rev. B \textbf{72}, 140503 (2005)
\bibitem{Yoko05b} T. Yokoyama, Y. Tanaka, A. A. Golubov, Y. Asano,
        Phys. Rev. B \textbf{72}, 214513 (2005)
\bibitem{Asul04} I. Asulin, A. Sharoni, O. Yulli, G. Koren, O. Millo,
        Phys. Rev. Lett. \textbf{93}, 157001 (2004)
\bibitem{Shar04} A. Sharoni, I. Asulin, G. Koren, O. Millo,
        Phys. Rev. Lett. \textbf{92}, 017003 (2004)
\bibitem{Pan00} S. H. Pan, E. W. Hudson, K. M. Lang, H. Eisaki,
        S. Uchida, J. C. Davis, 
        Nature \textbf{403}, 746 (2000)
\bibitem{Lang02} K. M. Lang, V. Madhavan, J. E. Hoffman, E. W. Hudson,
        H. Eisaki, S. Uchida, J. C. Davis, 
        Nature \textbf{415}, 412 (2002)
\bibitem{McEl03} K. McElroy, R. W. Simmonds, J. E. Hoffman, D.-H. Lee,
        J. Orenstein, H. Eisaki, S. Uchida, J. C. Davis, 
        Nature \textbf{422}, 592 (2003)
\bibitem{Vers04} M. Vershinin, S. Misra, S. Ono, Y. Abe, Y. Ando,
	A. Yazdani,
        Science \textbf{303}, 1995 (2004)
\bibitem{Hana04} T. Hanaguri, C. Lupien, Y. Kohsaka, D.-H. Lee,
	M. Azuma, M. Takano, H. Takagi, J. C. Davis,
        Nature \textbf{430}, 1001 (2004)
\bibitem{McEl05} K. McElroy, J. Lee, J. A. Slezak, D.-H. Lee, H. Eisaki,
        S. Uchida, J. C. Davis,
        Science \textbf{309}, 1048 (2005)
\bibitem{Apri98} M. Aprili, M. Covington, E. Paraoanu, B. Niedermeier, 
        L. H. Greene, 
        Phys. Rev. B \textbf{57}, R8139 (1998)
\bibitem{Apri99} M. Aprili, E. Badica, L. H. Greene, 
        Phys. Rev. Lett. \textbf{83}, 4630 (1999)
\bibitem{Wang99} W. Wang, M. Yamazaki, K. Lee, I. Iguchi, 
        Phys. Rev. B \textbf{60}, 4272 (1999)
\bibitem{Shar02} A. Sharoni, O. Millo, A. Kohen, Y. Dagan, R. Beck,
        G. Deutscher, G. Koren, 
        Phys. Rev. B \textbf{65}, 134526 (2002)
\bibitem{Iguc00} I. Iguchi, W. Wang, M. Yamazaki, Y. Tanaka, 
        S. Kashiwaya, 
        Phys. Rev. B \textbf{62}, R6131 (2000)
\bibitem{Sato96} M. Sato, 
        Physica C \textbf{263}, 271 (1996)
\bibitem{Idem90} Y. Idemoto, K. Fueki, 
        Jpn. J. Appl. Phys. \textbf{29}, 2729 (1990)
\bibitem{Yama90} A. Yamamoto, M. Onoda, E. Takayama-Muromachi, F. Izumi,
        T. Ishigaki, H. Asano, 
        Phys. Rev. B \textbf{42}, 4228 (1990)
\bibitem{Shig00} I. Shigeta, T. Uchida, Y. Tominari, T. Arai,
        F. Ichikawa, T. Fukami, T. Aomine, V. M. Svistunov, 
        J. Phys. Soc. Jpn. \textbf{69}, 2743 (2000)
\bibitem{Sinh98a} S. Sinha, K.-W. Ng, 
        Phys. Rev. Lett. \textbf{80}, 1296 (1998)
\bibitem{Sinh98b} S. Sinha, K.-W. Ng, 
        J. Phys. Chem. Solids \textbf{59}, 2078 (1998)
\bibitem{Covi97} M. Covington, M. Aprili, E. Paraoanu, L. H. Greene,
        F. Hu, J. Zhu, C. A. Mirkin,
        Phys. Rev. Lett. \textbf{79}, 277 (1997)
\bibitem{Tana02} Y Tanaka, Y. Tanuma, K. Kuroki, S. Kashiwaya, 
        J. Phys. Soc. Jpn. \textbf{71}, 2102 (2002)
\bibitem{Kita03} N. Kitaura, H. Itoh, Y. Asano, Y. Tanaka, J. Inoue,
        Y. Tanuma, S. Kashiwaya, 
        J. Phys. Soc. Jpn. \textbf{72}, 1718 (2003)
\bibitem{Dyne78} R. C. Dynes, V. Narayanamurti, J. P. Garno, 
        Phys. Rev. Lett. \textbf{41}, 1509 (1978)
\bibitem{Dyne84} R. C. Dynes, J. P. Garno, G. B. Hertel, T. P. Orlando, 
        Phys. Rev. Lett. \textbf{53}, 2437 (1984)
\bibitem{Suzu94} M. Suzuki, K. Komorita, M. Nagano, 
        J. Phys. Soc. Jpn. \textbf{63}, 1449 (1994)
\bibitem{Tana96} S. Tanaka, E. Ueda, M. Sato, K. Tamasaku, S. Uchida, 
        J. Phys. Soc. Jpn. \textbf{65}, 2212 (1996)
\bibitem{Alff97} L. Alff, H. Takashima, S. Kashiwaya, N. Terada,
	H. Ihara, Y. Tanaka, M. Koyanagi, K. Kajimura, 
        Phys. Rev. B \textbf{55}, R14757 (1997)
\bibitem{Kash98} S. Kashiwaya, T. Ito, K. Oka, S. Ueno, H. Takashima, 
        M. Koyanagi, Y. Tanaka, K. Kajimura, 
        Phys. Rev. B \textbf{57}, 8680 (1998)
\bibitem{Cren00} T. Cren, D. Roditchev, W. Sacks, J. Klein,
        Europhys. Lett. \textbf{52}, 203 (2000)
\bibitem{Gonn01} R. S. Gonnelli, A. Calzolari, D. Daghero, L. Natale,
	G. A. Ummarino, V. A. Stepanov, M. Ferretti,
        Eur. Phys. J. B \textbf{22}, 411 (2001)
\bibitem{Imai02} T. Imaizumi, T. Kawai, T. Uchiyama, I. Iguchi, 
        Phys. Rev. Lett. \textbf{89}, 017005 (2002)
\bibitem{Schm02} H. Schmidt, J. F. Zasadzinski, K. E. Gray, D. G. Hinks,
        Phys. Rev. Lett. \textbf{88}, 127002 (2002)
\bibitem{Kohe03} A. Kohen, G. Leibovitch, G. Deutscher,
        Phys. Rev. Lett. \textbf{90}, 207005 (2003)
\bibitem{Hoog03} B. W. Hoogenboom, C. Berthod, M. Peter, \O. Fischer,
        A. A. Kordyuk,
        Phys. Rev. B \textbf{67}, 224502 (2003)
\bibitem{Taka04} T. Takasaki, T. Ekino, T. Muranaka, T. Ichikawa,
	H. Fujii, J. Akimitsu,
        J. Phys. Soc. Jpn. \textbf{73}, 1902 (2004)
\bibitem{Ding96} H. Ding, M. R. Norman, J. C. Campuzano, M. Randeria,
	A. F. Bellman, T. Yokoya, T. Takahashi, T. Mochiku, K. Kadowaki,
        Phys. Rev. B \textbf{54}, R9678 (1996-II)
\bibitem{Itoh94} Y. Itoh, K. Yoshimura, T. Ohomura, H. Yasuoka, Y. Ueda,
	K. Kosuge, 
        J. Phys. Soc. Jpn. \textbf{63}, 1455 (1994)
\bibitem{Bour96} P. Bourges, L. P. Regnault, Y. Sidis, C. Vettier, 
        Phys. Rev. B \textbf{53}, 876 (1996-II)
\bibitem{Pana98} C. Panagopoulos, T. Xiang, 
        Phys. Rev. Lett. \textbf{81}, 2336 (1998)
\bibitem{Woll93} D. A. Wollman, D. J. Van Harlingen, W. C. Lee,
	D. M. Ginsberg, A. J. Leggett, 
        Phys. Rev. Lett. \textbf{71}, 2134 (1993)
\bibitem{Petr93} V. T. Petrashov, V. N. Antonov, P. Delsing, R. Claeson,
        Phys. Rev. Lett. \textbf{70}, 347 (1993)
\bibitem{Petr95} V. T. Petrashov, V. N. Antonov, P. Delsing, T. Claeson,
        Phys. Rev. Lett. \textbf{74}, 5268 (1995)
\bibitem{Wees92} B. J. van Wees, P. de Vries, P. Magn\'{e}e,
	T. M. Klapwijk,
Phys. Rev. Lett. \textbf{69}, 510 (1992)
\bibitem{Huan89} Q. Huang, J. F. Zasadzinski, K. E. Gray, J. Z. Liu,
        H. Claus, 
        Phys. Rev. B \textbf{40}, 9366 (1989)
\bibitem{Miya99} N. Miyakawa, J. F. Zasadzinski, L. Ozyuzer,
        P. Guptasarma, D. G. Hinks, C. Kendziora, K. E. Gray, 
        Phys. Rev. Lett. \textbf{83}, 1018 (1999)
\bibitem{Kita96} K. Kitazawa, H. Sugawara, T. Hasegawa, 
        Physica C \textbf{263}, 214 (1996)
\bibitem{Kash95} S. Kashiwaya,Y. Tanaka, M. Koyanagi, H. Takashima,
	K. Kajimura, 
        Phys. Rev. B \textbf{51}, 1350 (1995); 
        S. Kashiwaya, Y. Tanaka, M. Koyanagi, K. Kajimura,
        Phys. Rev. B \textbf{53}, 2667 (1996)

\end{thebibliography}
%

\end{document}